\documentclass[a4paper,12pt,english,reqno]{amsart}
\usepackage[T1]{fontenc}
\usepackage[latin9]{inputenc}
\usepackage[letterpaper]{geometry}
\usepackage{color}
\usepackage{amsthm}
\usepackage{amsmath}
\usepackage{graphicx}
\usepackage{setspace}
\usepackage{amssymb}
\usepackage[round]{natbib}
\usepackage[colorlinks=true,citecolor=blue,linkcolor=blue]{hyperref}

\makeatletter
\numberwithin{equation}{section}
\numberwithin{figure}{section}

\usepackage{graphicx}
\usepackage{stmaryrd}
\usepackage{amsmath, amstext,amsbsy,amsopn,amsthm, amscd,amsfonts,amssymb}

\usepackage{times}
\usepackage{mathptmx}

\makeatother

\usepackage{babel}

\global\long\def\dg{\mathbf{F}}
\global\long\def\dgcomp#1{F_{#1}}
\global\long\def\piola{\mathbf{P}}
\global\long\def\refbody{\Omega_{0}}
\global\long\def\refbnd{\partial\refbody}
\global\long\def\bnd{\partial\Omega}
\global\long\def\rcg{\mathbf{C}}
\global\long\def\lcg{\mathbf{b}}

\global\long\def\cronck#1{\delta_{#1}}
\global\long\def\lcgcomp#1{b_{#1}}
\global\long\def\deformation{\boldsymbol{\chi}}
\global\long\def\dgt{\dg^{\mathrm{T}}}
\global\long\def\idgcomp#1{F_{#1}^{-1}}
\global\long\def\velocity{\mathbf{v}}
\global\long\def\accel{\mathbf{a}}

\global\long\def\idg{\dg^{-1}}
\global\long\def\cauchycomp#1{\sigma_{#1}}
\global\long\def\idgt{\dg^{\mathrm{-T}}}
\global\long\def\cauchy{\boldsymbol{\sigma}}
\global\long\def\normal{\mathbf{n}}
\global\long\def\normall{\mathbf{N}}
\global\long\def\traction{\mathbf{t}}

\global\long\def\ed{\mathbf{d}}
\global\long\def\edcomp#1{d_{#1}}
\global\long\def\edl{\mathbf{D}}
\global\long\def\edlcomp#1{D_{#1}}
\global\long\def\ef{\mathbf{e}}
\global\long\def\efcomp#1{e_{#1}}
\global\long\def\efl{\mathbf{E}}

\global\long\def\surfacech{w_{e}}
\global\long\def\outer#1{#1^{\star}}
\global\long\def\perm{\epsilon_{0}}
\global\long\def\matper{\epsilon}
\global\long\def\jump#1{\left[\left[#1\right]\right]}
\global\long\def\identity{\mathbf{I}}
\global\long\def\area{\mathrm{d}a}
\global\long\def\areal{\mathrm{d}A}
\global\long\def\refsys{\mathbf{X}}
\global\long\def\Grad{\nabla_{\refsys}}
\global\long\def\grad{\nabla}
\global\long\def\divg{\nabla\cdot}
\global\long\def\Div{\nabla_{\refsys}}
\global\long\def\derivative#1#2{\frac{\partial#1}{\partial#2}}
\global\long\def\aef{\Psi}
\global\long\def\dltendl{\edl\otimes\edl}
\global\long\def\tr#1{\mathrm{tr}\left(#1\right)}
\global\long\def\ii#1{I_{#1}}
\global\long\def\dh{\hat{D}}
\global\long\def\inc#1{\dot{#1}}
\global\long\def\sys{\mathbf{x}}

\global\long\def\Curl{\nabla_{\refsys}}
\global\long\def\piolaincpush{\boldsymbol{\Sigma}}
\global\long\def\piolaincpushcomp#1{\Sigma_{#1}}
\global\long\def\edlincpush{\check{\mathbf{d}}}
\global\long\def\edlincpushcomp#1{\check{d}_{#1}}
\global\long\def\efincpush{\check{\mathbf{e}}}
\global\long\def\efincpushcomp#1{\check{e}_{#1}}
\global\long\def\elaspush{\boldsymbol{\mathcal{C}}}
\global\long\def\elecpush{\boldsymbol{\mathcal{A}}}
\global\long\def\elaselecpush{\boldsymbol{\mathcal{B}}}
\global\long\def\disgrad{\mathbf{h}}
\global\long\def\disgradcomp#1{h_{#1}}
\global\long\def\trans#1{#1^{\mathrm{T}}}
\global\long\def\elecpushcomp#1{\mathcal{A}_{#1}}
\global\long\def\elaselecpushcomp#1{\mathcal{B}_{#1}}
\global\long\def\elaspushcomp#1{\mathcal{C}_{#1}}
\global\long\def\dnh{\aef_{DH}}
\global\long\def\dnhc{\mu\lambda^{2}}
\global\long\def\dnhcc{\frac{\mu}{\lambda^{2}}+\frac{1}{\matper}d_{2}^{2}}
\global\long\def\dnhb{\frac{1}{\matper}d_{2}}
\global\long\def\afreq{\omega}
\global\long\def\dispot{\phi}
\global\long\def\edpot{\varphi}
\global\long\def\newcofa{A_{1}^{+}}
\global\long\def\newcofaa{A_{5}^{+}}
\global\long\def\newcofb{B^{-}}
\global\long\def\newcofc{C^{+}}

\global\long\def\kh{\hat{k}}
\global\long\def\afreqh{\hat{\afreq}}
\global\long\def\phasespeed{c}
\global\long\def\bulkspeed{c_{B}}
\global\long\def\speedh{\hat{c}}
\global\long\def\dhth{\dh_{th}}

\global\long\def\maxinccomp#1{\inc{\outer{\sigma}}_{#1}}
\global\long\def\maxcomp#1{\outer{\sigma}_{#1}}
\global\long\def\relper{\matper_{r}}
\global\long\def\sdh{\hat{d}}
\global\long\def\sdhth{\sdh_{th}}

\global\long\def\kcos{\kh_{co}^{sym}}
\global\long\def\kcoa{\kh_{co}^{anti}}

\begin{document}

\title{The Rayleigh-Lamb wave propagation in dielectric elastomer layers
subjected to large deformations}

\author{Gal  Shmuel{$^{\dagger}$}, Massimiliano Gei {$^{\ddagger}$} and Gal deBotton {$^{\dagger,\S}$}\\ \\
\small {$^{\dagger}$} The Pearlstone Center for Aeronautical Studies,\\
\small Department of Mechanical Engineering,\\
\small Ben-Gurion University,\\
\small Beer-Sheva 84105, Israel.\\
\\
\small {$^{\ddagger}$} Department of Mechanical and Structural Engineering,\\
\small University of Trento,\\
\small via Mesiano 77, 38123 Trento, Italy.\\ 
\\
\small {$^{\S}$} Department of Biomedical Engineering,\\
\small Ben-Gurion University,\\
\small Beer-Sheva 84105, Israel.}

\maketitle

\begin{abstract}
The propagation of waves in soft dielectric elastomer layers is investigated. To this end incremental motions superimposed on homogeneous finite deformations induced by bias electric fields and pre-stretch are determined. First we examine the case of mechanically traction-free layer, which is an extension of the Rayleigh-Lamb problem in the purely elastic case. Two other loading configurations are accounted for too. Subsequently, numerical examples for the dispersion relations are evaluated for a dielectric solid governed by an augmented neo-Hookean strain energy.
It is found that the the phase speeds and frequencies strongly depend on the electric excitation and pre-stretch. These findings lend themselves at the possibility of controlling the propagation velocity as well as filtering particular frequencies with suitable choices of the electric bias field.
\end{abstract}

\section{Introduction}

The goal of this work is to investigate the  propagation of electromechanical induced waves in a dielectric
elastomer (DE) layer subjected to finite deformations. In the purely elastic case the first solution for
surface waves based on the exact equations of (2D) elasticity was introduced
by Lord \citeauthor{rayleigh87} in 1887, who determined  the so-called
Rayleigh waves. This work was later extended for propagation of
waves in elastic plates by Lord \citet{lordrayleigh89} himself and \citet{lamb89}.
Here we extend  the Rayleigh-Lamb problem and
account for incremental motions superimposed on finite deformations
in dielectric media, and  investigate how these are influenced
by the presence of  external electric field and pre-stretch.

When subjected to an electric field electroactive polymers (EAPs)
deform and both their mechanical and electrical properties are modified.
In contrast to piezoelectric ceramics, DEs are capable of undergoing
large deformations, a property that entitled them the name \textquotedblleft{}artificial
muscles\textquotedblright{}. Moreover, while in piezoelectricity the
electromechanical coupling is linear, in DEs the mechanical field depends quadratically on the   electric
fields. A proper theory, which accounts
for the aforementioned coupling
and captures the ability of the material to undergo finite strains
is therefore required. The foundations of this nonlinear electroelastic
theory are summarized in the pioneering works of \citet{toup56arma}
and \citet{eringen63engsci} for the static case. These contributions
were later extended by \citet{Toupin1963a} to account for the dynamics
of these elastic dielectrics. A comprehensive summary can be found in monographs by \citet{eringenbook90}
and \citet{kovetz00}. 
Due to the development of new materials
that admit this coupled behavior, thus branching toward a window of
new applications (e.g., \citealp{pelr&etal00scie}, \citealp{barc02js&r}
and \citealp{caprismela09}), the interest in these electroelastodynamic
theories revived recently and the coupled electromechanical theory was revisited recently \citep[e.g.,][]{mcme&land05jamt,dorf&ogde05acmc,eric07arma,suo&etal08jmps}.
The foregoing works and their extension
to the domain of soft dielectric composites by \citet{gdb&etal07mams}
and \citet{Bertoldi2011}, differ in their constitutive formulations,
the choice of the independent variables, the resultant electrostatic
stress-like tensors and electric body-like forces. For a
review of the diverse formulations of the constitutive laws and governing
equations the reader is referred to \citet{Bustamante2009}. Among
the various approaches we recall the formulation proposed by  \citet{dorf&ogde05acmc}
for the static case, and its extension to dynamics  by  \citet{Dorfmannogden2010}.
In these works  the concept of 'total' stress tensor that is derived from a
 'total' or 'augmented' energy-density function was employed. In this work we follow the framework proposed by the latter.

In contrast with the relatively large body of theoretical works that
are available, only a few boundary-value problems (BVP) were solved 
in the context of the  dynamic behavior of  electromechanically coupled  
EAPs. One of the first contributions in the field of
piezoelectricity was made by \citet{Tiersten1963}, who examined  the
 thickness vibrations of an infinite piezoelectric plate
induced by alternating voltage at the  surface electrodes, and later
solved the corresponding problem of wave propagation  \citep{Tiersten1963a}.
To the best of the authors knowledge, while for piezoelectric media
solutions for BVP accounting for the influence of pre-strain and bias
field are available (see review article by \citealp{Yang2004}),
analogue developments in the context of dielectrics were considered by \citet{mock&goul06ijnm} and \citet{Zhu2010}, who examined the dynamic behavior of dielectric elastomer balloons, and by \citet{Dorfmannogden2010}
who studied the problem of propagation of Rayleigh surface waves in
a dielectric half-space. 
Herein we continue along the path of the latter contribution and consider the extension of the Rayleigh-Lamb wave propagation problem to finitely deformed dielectric layers subjected to coupled electromechanical loading.

The work is composed as follows. In section 2 the theory of finite
electroelastodynamics is summarized. The equations for incremental
motions superimposed on finite deformations are outlined  in section
3. Specific finite deformations that correspond to three different
loading configurations are considered in section 4 for a layer whose
behavior is characterized by a particular augmented energy-density
function (AEDF), namely the \emph{incompressible dielectric neo-Hookean}
model (DH). Next, the extension of the Rayleigh-Lamb dispersion relation
for a DH layer is introduced in section 5. An analysis of the
dispersion relation is carried out in section 6, and the effects of
the bias electrostatic field and pre-stretch are investigated for
 the three loading configurations. The main conclusions and
observations are summarized in section 7.

\section{Finite electroelasticity}

Let $\deformation:\refbody\times\mathcal{I}\rightarrow\Omega\subset\mathbb{R}^{3}$
describe the motion of a material point $\mathbf{X}$ from a reference
configuration of a body $\Omega_{0}$, with a boundary $\refbnd$,
to its current configuration $\Omega$, with boundary $\bnd$, by $\sys=\deformation\left(\mathbf{X},t\right)$, where $\mathcal{I}$ is a time interval. The domain of the space surrounding
the body is $\mathbb{R}^{3}\backslash\Omega$ and is assumed to be
vacuum. The corresponding velocity and acceleration are denoted by
$\velocity=\deformation_{,t}$ and $\accel=\deformation_{,tt},$ respectively,
while the deformation gradient is $\dg=\frac{\boldsymbol{\partial}\deformation}{\boldsymbol{\partial}\mathbf{\refsys}}=\Grad\deformation$,
and where due to the material impenetrability $J\equiv\det\left(\mathbf{F}\right)>0$.
Vectors between two infinitesimally close points are related through
$\mathrm{d}\mathbf{x}=\dg\mathrm{d}\mathbf{X}$, whereas area elements
are transformed via Nanson's formula $\normall\areal=\frac{1}{J}\dgt\normal\area$.
The volume ratio between an infinitesimal volume element $\mathrm{d}v$
in the deformed configuration, and its counterpart in the reference
$\mathrm{d}V$ is given by $\mathrm{d}v=J\mathrm{d}V$. As measures
of the deformation the right and left Cauchy-Green strain tensors
$\rcg=\dgt\mathbf{F}$ and $\lcg=\dg\dgt$ are used.

Let $\ef$ denote the electric field in the current configuration.
Commonly the electric field is given by means of a gradient of a scalar
field, namely the electrostatic potential. The induced electric displacement
field $\ed$ is related to the electric field in free space via the
vacuum permittivity $\perm$ such that $\ed=\perm\ef$. In dielectric
media an appropriate constitutive law specifies the relationship between
these fields. Generally, this connection can be non-linear and anisotropic.

The balance of linear momentum is \begin{equation}
\divg\cauchy=\rho\accel,\label{eq:eom}\end{equation}
where $\cauchy$ is the 'total' stress tensor, and $\rho$ is the
material mass density. The balance of angular momentum implies that
$\cauchy$ is symmetric. Note that $\cauchy$ consists of both mechanical
and electrical contributions, such that the traction $\traction$ on a deformed  area element can be written as  $\cauchy\normal$ where $\normal$ is unit vector normal to $\bnd$. 
On the boundary of the material we postulate a separation of the traction into the sum
of a mechanical traction $\traction_{m}$ which is a prescribed data, and an electrical traction $\traction_{e}$ which is induced by the external electric field.

Assuming no free body charge (ideal dielectric), Gauss'  law reads\begin{equation}
\divg\ed=0.\label{eq:gauss law}\end{equation}
Under a quasi-electrostatic approximation, appropriate when for the same frequency the length of the waves under consideration are shorter than the electromagnetic waves,  Faraday's
law states that the electric field is curl-free, i.e., \begin{equation}
\grad\times\ef=\mathbf{0},\label{eq:faraday law}\end{equation}
thus enabling the usage of the aforementioned   electrostatic potential.

Taking into account fields outside   the material, which henceforth
will be identified by a star superscript, the following jump conditions should be
satisfied across   $\bnd$, namely

\begin{eqnarray}
\jump{\cauchy}\normal=\traction_{m},\quad & \jump{\ed}\cdot\normal=-\surfacech, & \quad\jump{\ef}\times\normal=\mathbf{0},\label{eq:jumps}\end{eqnarray}
where  $\surfacech$ is the surface charge density, and the notation
$\jump{\bullet}=\left(\bullet\right)-\outer{\left(\bullet\right)}$
is used for the difference between fields inside and outside of the
material. The outer fields are related by\begin{eqnarray}
\outer{\ed} & = & \perm\outer{\ef},\label{eq:outer d e}\\
\outer{\cauchy} & = & \perm\left[\outer{\ef}\otimes\outer{\ef}-\frac{1}{2}\left(\outer{\ef}\cdot\outer{\ef}\right)\identity\right],\label{eq:maxwell stress}\end{eqnarray}
where $\identity$ is the identity tensor. Herein we identify the electrical traction $\traction_{e}$ as the consequence  of the external stress $\outer{\cauchy} $, namely the Maxwell stress,  such that $\traction_{e}=\outer{\cauchy}\normal$. In the surrounding space outside
the material $\outer{\ed}$ and $\outer{\ef}$ must satisfy Eqs.
\eqref{eq:gauss law}-\eqref{eq:faraday law}, which reduce to Laplace
equation of the electrostatic potential. As a consequence the Maxwell
stress is divergence-free.

The foregoing balance and jump equations can be recast in a Lagrangian
formulation with the appropriate \emph{pull-back} operations. Specifically,
we have that \begin{eqnarray}
\piola=J\cauchy\idgt, & \quad\edl=J\idg\ed, & \quad\efl=\dgt\ef,\label{eq:pull back p d e}\end{eqnarray}
for the 'total' first Piola-Kirchhoff stress, Lagrangian electric
displacement and electric field, respectively (e.g., \citealp{dorf&ogde05acmc}). The corresponding balance
equations are \begin{eqnarray}
\Div\cdot\piola=\rho_{L}\accel, & \quad\Div\cdot\edl=0, & \quad\Curl\times\efl=\mathbf{0},\label{eq:l eom d e}\end{eqnarray}
where $\rho_{L}=J\rho$ is the density of the material in the reference
configuration. The jump conditions across the boundary $\refbnd$
read
\begin{eqnarray}
\jump{\piola}\normall=\traction_{M}, & \quad\jump{\edl}\cdot\normall=-w_{E}, & \quad\jump{\efl}\times\normall=\mathbf{0},\label{eq:lagrng jumps}
\end{eqnarray}
where $\traction_{M}\areal=\traction_{m}\area$, $w_{E}\areal=\surfacech\area$ and $\normall$ is a unit outward normal to $\refbnd$.

Following \citet{dorf&ogde05acmc}, the 'total' first Piola-Kirchhoff
stress and the Lagrangian electric field are given in terms of an
\emph{augmented} energy-density function $\aef$\emph{ }(AEDF) with
the independent variables $\dg$ and $\edl$, such that \begin{equation}
\piola=\derivative{\aef}{\dg},\quad\efl=\derivative{\aef}{\edl}.\label{eq:constitutive law}\end{equation}
For an incompressible material a Lagrange multiplier $p$ is introduced,
which is a workless reaction to the kinematic constraint such that
\begin{equation}
\piola=\derivative{\aef}{\dg}-p\idgt.\label{eq:stress for inc}\end{equation}
The latter can be determined only from the equilibrium equations and
the boundary conditions.

\section{Small fields superimposed on finite deformations}

Recent experiments reveal  how pre-stretching of dielectric elastomers
enhances properties such as the actuation strain (\citealp{pelr&etal00scie})
and breakdown strength (\citealp{Plante2006}). A thorough investigation
of the subject was recently done by \citet{kofod08}. Motivated by
these findings, we address the response of pre-stretched dielectric
elastomers to incremental deformations.

Following the formulation given in \citet{Dorfmannogden2010}, we
consider infinitesimal time-dependent elastic displacement and electric displacement increments
$\inc{\sys}=\inc{\deformation}\left(\refsys,t\right)$ and $\inc{\edl}\left(\refsys,t\right)$,
respectively, superimposed on the foregoing static configuration $\Omega$
reached via $\deformation\left(\mathbf{X}\right)$. Herein and throughout
this work a superposed dot will denote the incremental quantities.
The corresponding balance laws can be formulated in terms of  Eulerian
quantities, namely \begin{eqnarray}
\divg\piolaincpush=\rho\inc{\sys}_{,tt}, & \quad\divg\edlincpush=0, & \quad\grad\times\efincpush=\mathbf{0},\label{eq:inc eom gauss farady current conf}\end{eqnarray}
such that $\piolaincpush\ \,\edlincpush\ \mathrm{and}\ \efincpush$
are, respectively, the \emph{push-forwards} of the increments in the total
first Piola-Kirchhoff stress, the Lagrangian electric displacement and
the electric field determined via the inverse of the transformations given
in Eq.~\eqref{eq:pull back p d e}. Next, upon linearization, the
incremental constitutive equations for an incompressible material
are\begin{eqnarray}
\piolaincpush & = & \elaspush\disgrad+p\disgrad^{\mathrm{T}}-\inc p\identity+\elaselecpush\edlincpush,\label{eq:constitutive eq for the push of piola}\\
\efincpush & = & \elaselecpush^{\mathrm{T}}\disgrad+\elecpush\edlincpush,\label{eq:constitutive eq for the push of inc E}\end{eqnarray}
where $\left(\elaselecpush^{\mathrm{T}}\disgrad\right)_{k}=\mathcal{B}_{ijk}h_{ij}$. For a compressible
material Eq.~\eqref{eq:constitutive eq for the push of piola} is
to be taken with $p\equiv\dot{p}\equiv0$. Herein $\disgrad=\grad\inc{\sys}$
is the displacement gradient and \begin{eqnarray}
\elecpushcomp{ij}=J\idgcomp{\alpha i}\idgcomp{\beta j}\elecpushcomp{0\alpha\beta}, & \quad\elaselecpushcomp{ijk}=\dgcomp{j\alpha}\idgcomp{\beta k}\elaselecpushcomp{0i\alpha\beta}, & \quad\elaspushcomp{ijkl}=\frac{1}{J}\dgcomp{j\alpha}\dgcomp{l\beta}\elaspushcomp{0i\alpha k\beta},\label{eq:A B C push def}\end{eqnarray}
are the push-forwards of the referential electric, electroelastic,
and elasticity tensors, respectively. The latter are defined by\begin{eqnarray}
\elecpushcomp{0\alpha\beta}=\derivative{^{2}\aef}{\edlcomp{\alpha}\partial\edlcomp{\beta}},\quad\quad\elaselecpushcomp{0i\alpha\beta}=\derivative{^{2}\aef}{\dgcomp{i\alpha}\partial\edlcomp{\beta}}, &  & \quad\elaspushcomp{0i\alpha k\beta}=\derivative{^{2}\aef}{\dgcomp{i\alpha}\partial\dgcomp{k\beta}}.\label{eq: A0 B0 C0 comp}\end{eqnarray}
Similarly, the incremental outer fields are \begin{eqnarray}
\inc{\outer{\ed}} & = & \perm\inc{\outer{\ef}},\label{eq:in electro outer fields}\\
\inc{\outer{\cauchy}} & = & \perm\left[\inc{\outer{\ef}}\otimes\outer{\ef}+\outer{\ef}\otimes\inc{\outer{\ef}}-\left(\outer{\ef}\cdot\inc{\outer{\ef}}\right)\identity\right],\label{eq:inc maxwell stress}\end{eqnarray}
where $\inc{\outer{\ed}}$ and $\inc{\outer{\ef}}$ are to satisfy
Eqs.~\eqref{eq:gauss law}-\eqref{eq:faraday law}, hence $\inc{\outer{\cauchy}}$
is identically divergence-free.

Utilizing several kinematic relations (see \citealp{Dorfmannogden2010})
the corresponding jump conditions are \begin{eqnarray}
\left[\piolaincpush-\inc{\outer{\cauchy}}+\outer{\cauchy}\trans{\disgrad}-\left(\divg\inc{\sys}\right)\outer{\cauchy}\right]\normal & = & \check{\traction}_{m},\label{eq:in stress jump}\\
\left[\edlincpush-\inc{\outer{\ed}}-\left(\divg\inc{\sys}\right)\outer{\ed}+\disgrad\outer{\ed}\right]\cdot\normal & = & -\check{w}_{e},\label{eq:in charge  jump}\\
\left[\mathrm{\efincpush}-\inc{\outer{\ef}}-\trans{\disgrad}\outer{\ef}\right]\times\normal & = & \mathbf{0},\label{eq:inc elec jump}\end{eqnarray}
where the quantities $\check{\traction}_{m}$ and $\check{w}_{e}$
are defined by $\check{\traction}_{m}\area=\dot{\traction}_{M}\areal$
and $\check{w}_{e}\area=\dot{w}_{E}\areal$.

\section{Incompressible dielectric neo-Hookean layer subjected to transverse
bias field}
We recall that the constitutive laws in terms of the AEDF has the deformation gradient and the electric displacement field as its independent variables. Whenever the material is isotropic its symmetry can be exploited,
and the dependency of the AEDF on $\dg$ and $\edl$ is reduced to
the appropriate invariants of $\rcg$ and $\dltendl$, namely \begin{equation}
\begin{array}{lcc}
I_{1}=\tr{\rcg}=\rcg:\mathbf{I}, & I_{2}=\frac{1}{2}\left(I_{1}^{2}-\rcg:\rcg\right), & I_{3}=\det\left(\rcg\right)=J^{2},\end{array}\label{eq:c invariants}\end{equation}
and\begin{eqnarray}
\ii{4e}=\tr{\dltendl}, & \ii{5e}=\rcg:\left(\dltendl\right), & \ii{6e}=\rcg^{2}:\left(\dltendl\right).\label{eq:dd invariants}\end{eqnarray}

Thus, we consider the AEDF \begin{equation}
\dnh\left(\ii 1,\ii{5e}\right)=\frac{\mu}{2}\left(\ii 1-3\right)+\frac{1}{2\matper}\ii{5e},\label{eq:dielecric neo hookean}\end{equation}
where $\mu$ is the shear modulus and $\matper$ is the material permittivity
that equals the vacuum permittivity $\perm$ times the material relative
permittivity $\relper$. This model recovers a neo-Hookean behavior
in a purely elastic case, as well as the simple isotropic linear relation
$\ed=\matper\ef$ between the current displacement and electric field.
Henceforth we refer to Eq.~\eqref{eq:dielecric neo hookean} as the
\emph{incompressible dielectric neo-Hookean} model (DH). Throughout
this work the material behavior is assumed to be governed by the DH
model. The corresponding total stress is \begin{equation}
\cauchy=\mu\lcg+\frac{1}{\matper}\ed\otimes\ed-p\identity,\label{eq:d nh stress}\end{equation}
where, in component form, the constitutive tensors $\elecpush,\ \elaselecpush,\ \mathrm{and\ }\elaspush$
are \begin{eqnarray}
\elecpushcomp{ij}=\frac{1}{\matper}\cronck{ij}, & \quad\elaselecpushcomp{ijk}=\frac{1}{\matper}\left(\cronck{ik}\edcomp j+\edcomp i\cronck{jk}\right), & \quad\elaspushcomp{ijkl}=\mu\cronck{ik}\lcgcomp{jl}+\frac{1}{\matper}\cronck{ik}\edcomp j\edcomp l.\label{eq:dnh a b  c comp}\end{eqnarray}

Consider a Cartesian coordinate system with unit vectors $\mathbf{i_{1},i_{2}}$
and $\mathbf{i_{3}}$ along the $x_{1},x_{2}$ and $x_{3}$ axes, respectively.
Let a layer governed  by the DH model be infinitely long along the $x_{1}$-axis  
with a thickness $2h$ along the $x_{2}$-axis. Assuming
a plane-strain configuration there is  no deformation along the
 $x_{3}$-axis and the fields being independent of $x_{3}$.
An electric displacement field is applied to the layer along its thickness.
In the current configuration this field is given by $\ed=\edcomp 2\mathbf{i_{2}}$,
and the corresponding electric field equals the voltage difference
between the electrodes divided by $2h$. The resultant deformation in terms of a homogeneous diagonal deformation gradient $\mathrm{F}=\mathrm{diag}\left[\lambda,\lambda^{-1},1\right]$,
is related to the stress via\begin{eqnarray}
\cauchycomp{11}=\mu\lambda^{2}-p, & \quad\cauchycomp{22}=\frac{\mu}{\lambda^{2}}+\frac{1}{\epsilon}\edcomp 2^{2}-p.\label{eq:stress components bias}\end{eqnarray}
Specialization of the in-plane components of Eq.~\eqref{eq: A0 B0 C0 comp}
for the considered deformation gives\begin{eqnarray}
\elecpushcomp{11} & = & \elecpushcomp{22}=\frac{1}{\matper},\\
\elaselecpushcomp{121} & = & \elaselecpushcomp{211}=\frac{1}{2}\elaselecpushcomp{222}=\frac{1}{\matper}\edcomp 2,\\
\elaspushcomp{1111} & = & \elaspushcomp{2121}=\mu\lambda^{2},\\
\elaspushcomp{1212} & = & \elaspushcomp{2222}=\frac{\mu}{\lambda^{2}}+\frac{1}{\epsilon}\edcomp 2^{2}.\end{eqnarray}
Wave propagation and vibrations of the layer are addressed
for  the following three fundamental
loading paths.

\subsection*{Path A: expansion-free finite loading. }

\begin{figure}[t]
\includegraphics[scale=0.30]{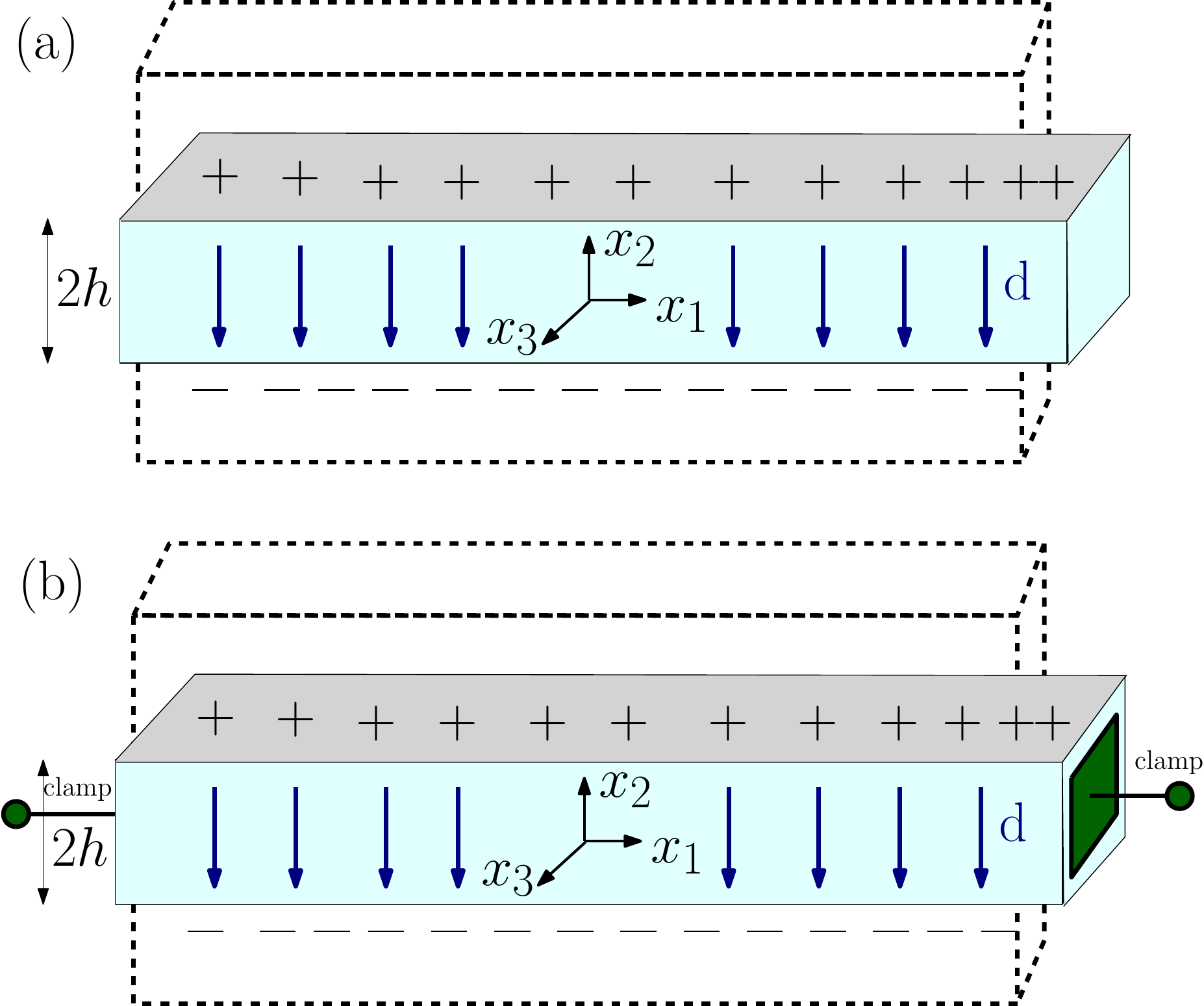} \includegraphics[scale=0.33]{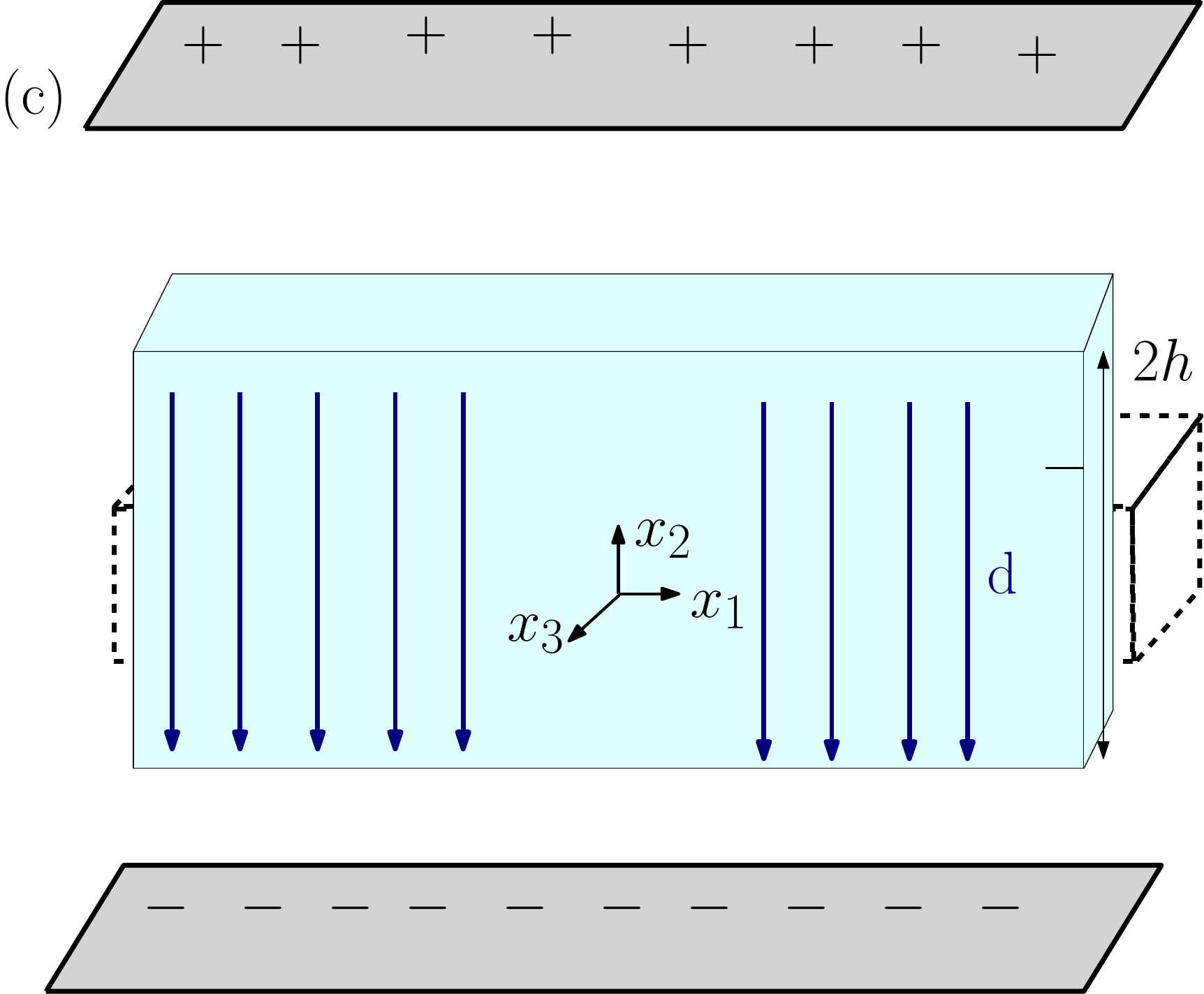}

\caption{Illustration of (a) path A: expansion-free loading, (b) path
B: initially pre-stretched layer, and (c) path C: a layer immersed
in electric field.}

\label{all paths}
\end{figure}

The layer having its top and bottom surfaces $x_{2}=\pm h$ coated
with soft electrodes and mechanically traction free ($\traction_{m}=\mathbf{0}$).
It is also free to expand along the $x_{1}$-axis (Fig.
\ref{all paths}a). From the traction-free boundary conditions
simple connections between the stretch, the pressure and the Lagrangian
electric displacement field $\edl=\lambda\ed$ are obtained,
namely
\begin{eqnarray}
\lambda=\left(1+\hat{D}^{2}\right)^{1/4}, & \quad & p=\mu\left(1+\dh^{2}\right)^{1/2},\label{eq:l d p d}\end{eqnarray}
where $\dh=\edlcomp 2/\sqrt{\mu\matper}$. Note that due to the symmetry of the problem the electric fields along with the Maxwell stress outside
the layer vanish. The surface charge $w_{E}$ is evaluated via Eq.
\eqref{eq:lagrng jumps}, recalling that $\outer{\ed}=\mathbf{0}$.

\subsection*{Path B: initially pre-stretched layer. }

Once again the surfaces $x_{2}=\pm h$ are coated with electrodes
and are free of mechanical traction. The layer is first pre-stretched
along the  $x_{1}$-axis to $\lambda=\tilde{\lambda}$,
and then clamped (Fig.~\ref{all paths}b). Subsequently, an electric
displacement field is applied, while the longitudinal stretch is kept constant. The pressure is given in terms of the pre-stretch
and the electric displacement field as \begin{equation}
p=\mu\left(\frac{1}{\tilde{\lambda}^{2}}+\sdh^{2}\right),\label{eq:p second path}\end{equation}
where $\sdh=\edcomp 2/\sqrt{\mu\matper}$. The resultant pre-stress
is obtained via Eq.~\eqref{eq:stress components bias}. Again,
the electric fields and the Maxwell stress outside the layer vanish.
The surface charge $w_{e}$ is evaluated via Eq.~\eqref{eq:jumps},
having that $\outer{\ed}=\mathbf{0}$.

\subsection*{Path C: a layer immersed in electric field. }

The layer is immersed in a pre-existing electric field. The surfaces are free
of mechanical traction. The absence of the electrodes on the surfaces
implies that $w_{E}=0$, where the  pressure and the stretch are determined
via the corresponding jump conditions to yield\textcolor{red}{{} }\begin{eqnarray}
\lambda=\left(1-\dh^{2}\left(\relper-1\right)\right)^{1/4}, & \quad p=\frac{\mu\dh^{2}}{2\lambda^{2}}\left(1+\lambda^{4}\right).\label{eq: lambda d third path}\end{eqnarray}
We note that when deriving the solution it was assumed the electric
field is homogeneous outside the layer, neglecting fringing effects.
Thus, this solution is by no means exact, and serves only as a qualitative
result.
Eq. (\ref{eq: lambda d third path}a) reveals that the layer contracts
along the  $x_{1}$-axis and expands along the direction of
the electric displacement, contrary to the situation in path A.

\section{Wave propagation in dielectric layers}

We determine next the mechanical and electrical waves due
to harmonic excitation along the  $x_{1}$-axis superimposed
on the deformed configurations described previously. Following \citet{Dorfmannogden2010}, the incremental
equations of motion (\ref{eq:inc eom gauss farady current conf}a)
read
\begin{align}
 & \left(\dnhc+p\right)\disgradcomp{11,1}-\inc p_{,1}+\left(\dnhcc\right)\disgradcomp{12,2}+p\disgradcomp{21,2}+\dnhb\edlincpushcomp{1,2}
 =\rho\inc x_{1,tt},
 \label{eq:eom x} \\
 & \left(\dnhc+p\right)\disgradcomp{21,1}+p\disgradcomp{12,1}+\dnhb\edlincpushcomp{1,1}+\left(\dnhcc+p\right)\disgradcomp{22,2}-\inc p_{,2}+2\dnhb\edlincpushcomp{2,2}
 =\rho\inc x_{2,tt},
 \label{eq:eom y}
 \end{align}
while the incremental Faraday's equation (\ref{eq:inc eom gauss farady current conf}c)
is\begin{equation}
\dnhb\left(\disgradcomp{12,2}+\disgradcomp{21,2}\right)+\frac{1}{\matper}\edlincpushcomp{1,2}-2\dnhb\disgradcomp{22,1}-\frac{1}{\matper}\edlincpushcomp{2,1}=0.\label{eq:parady exp}\end{equation}
The motion has to satisfy the incompressibility constraint and Gauss'
equation, namely \begin{eqnarray}
\disgradcomp{11}+\disgradcomp{22}=0,\quad & \edlincpushcomp{1,1}+\edlincpushcomp{2,2}=0,\label{eq:inc}\end{eqnarray}
which motivate  the use of stream functions $\dispot\left(x_{1},x_{2},t\right)$
and $\edpot\left(x_{1},x_{2},t\right)$ such that \begin{eqnarray}
\inc x_{1}=\dispot_{,2},\quad\inc x_{2}=-\dispot_{,1}, & \quad\edlincpushcomp 1=\edpot_{,2}, & \quad\edlincpushcomp 2=-\edpot_{,1}.\end{eqnarray}
Normalization by $\dnhc$, followed by differentiation of Eqs.~\eqref{eq:eom x} and \eqref{eq:eom y} with respect to $x_{2}$ and $x_{1}$, respectively, and then subtraction of the latter form the former
results in the coupled equation
\begin{equation}
 \dispot_{,1111}+\left(1+\frac{1+\dh^{2}}{\lambda^{4}}\right)\dispot_{,1122}+\frac{1+\dh^{2}}{\lambda^{4}}\dispot_{,2222}+\frac{\dh}{\lambda^{3}\sqrt{\mu\matper}}\left(\edpot_{,112}+\edpot_{,222}\right)\label{eq:eom 1}
 =\frac{1}{\lambda^{2}}\frac{\rho}{\mu}\left(\dispot_{,11}+\dispot_{,22}\right)_{,tt},
  \end{equation}
After multiplying Eq.~\eqref{eq:parady exp} by  $\lambda\sqrt{\frac{\matper}{\mu}}$
it may be rewritten in the form
\begin{eqnarray}
\dh\left(\dispot_{,112}+\dispot_{,222}\right)+\frac{\lambda}{\sqrt{\mu\matper}}\left(\edpot_{,11}+\edpot_{,22}\right) & = & 0.\label{eq:new farady}\end{eqnarray}
The last two equations constitute a system of two coupled equations
for the stream functions $\dispot$ and $\edpot$. A solution
is feasible once the current stretch $\lambda$ and the dimensionless
nominal electric displacement $\hat{\mathrm{D}}$ are prescribed  in addition
to the density $\rho$ and shear modulus $\mu$ of the layer. We recall
that while for the paths A and C the quantities $\hat{\mathrm{D}}$
and $\lambda$ are uniquely related via Eqs. (\ref{eq:l d p d}a)
and (\ref{eq: lambda d third path}a), respectively, they are independent
for path B. Also note that the quantity $1/\sqrt{\mu\matper}$
can be integrated into the amplitude of the stream function $\edpot$.

Next we assume a solution with  time-dependency
in the form $e^{-i\afreq t}$   and   a periodicity
along the $x_{1}$-axis in the form $e^{ikx_{1}}$, where
$\afreq$ is the angular frequency and $k$ is the associated wavenumber,
such that the wave velocity is $c=\afreq/k$. Thus, traveling waves
solutions in the form \begin{eqnarray}
\dispot=Ae^{kqx_{2}}e^{ik\left(x_{1}-\phasespeed t\right)}, & \quad\edpot=kBe^{kqx_{2}}e^{ik\left(x_{1}-\phasespeed t\right)},\label{eq:phi1,2}\end{eqnarray}
are sought, where the appropriate values of $q$ are to be found from
the governing equations. Insertion of Eq.~\eqref{eq:phi1,2} into
Eqs.~\eqref{eq:eom 1}-\eqref{eq:new farady} yields the following
system of two linear homogeneous equations in $A$ and \textbf{$B$,
}namely \begin{eqnarray}
\left(q^{2}-1\right)\left(\frac{\rho}{\mu}\phasespeed^{2}+q^{2}\frac{1+\dh^{2}}{\lambda^{2}}-\lambda^{2}\right)A+\left(q^{2}-1\right)q\frac{1}{\lambda}\dh\hat{B} & = & 0,\label{eq:sys a b 1}\\
q\left(q^{2}-1\right)\dh A+\left(q^{2}-1\right)\lambda\hat{B} & = & 0,\label{eq:sys a b 2}\end{eqnarray}
where $\hat{B}=B/\sqrt{\mu\matper}$.
A non-trivial solution is possible for values of $q$ for which the determinant of the coefficients of A and B vanish.
This leads to a bi-cubic equation in $q$. The six roots
include two repeated ones, namely $q_{1}=q_{2}=1$ and $q_{3}=q_{4}=-1$,
along with roots $q_{5}=-q_{6}=\lambda\left(\lambda^{2}-\frac{\rho}{\mu}\phasespeed^{2}\right)^{1/2}$.

An interesting case arises if $\phasespeed^{2}=\lambda^{2}\frac{\mu}{\rho}$,
that is the square of the speed of a bulk shear wave in an isotropic elastic medium
subjected to a finite deformation propagating along the axis with principal stretch  $\lambda$.
In this case $q_{5}=q_{6}=0$,
and the corresponding solution is associated with propagation of waves
parallel to the boundary along the $x_{1}$-axis. As observed
by  \citet{Dorfmannogden2010} for the  half-space case, the associated
stream functions vanish as expected since an in-plane bulk shear wave
cannot propagate in this manner.

Usually, the general solutions for $\dispot$ and $\edpot$ are obtained
as a linear combination of the  roots in the form of \begin{eqnarray}
\dispot=\sum_{n=1}^{6}A_{n}e^{kq_{n}x_{2}}e^{ik\left(x_{1}-\phasespeed t\right)}, &  & \edpot=\sum_{n=1}^{6}B_{n}e^{kq_{n}x_{2}}e^{ik\left(x_{1}-\phasespeed t\right)}.\end{eqnarray}
However, a somewhat different structure of the solution is required for
the DH model. For the general case when the constitutive law involves
$\ii{4e}$, a full coupling between the mechanical and electric equations
is expected (see \citealp{Dorfmannogden2010}), and hence dependency
between $A_{n}$ and $B_{n}$ exists for all $n$. However, for the
specific constitutive behavior considered here we find that for the
roots $\left\{ q_{n}\right\} $, $n=1,..,4$, the coefficients associated
with $A_{1},...,A_{4},\,B_{1},...,B_{4}$ in the counter diagonal of the
coefficients matrix vanishes. Mathematically, this implies that the constants
$A_{1},...,A_{4}$ and $B_{1},...,B_{4}$ are independent. 
Physically, this hints at the possibility of propagation of elastic waves without
excitation of the accompanying electric waves. 
Thus, the associated solutions for
$\dispot$ and $\edpot$ of these roots are not to be taken as multiple.
The coupling between the coefficients still holds when $q_{5}$ and $q_{6}$
are considered, giving a relation between $A_{5}$ and $B_{5},$
and between $A_{6}$ and $B_{6}$. This relation is obtained by substituting
$q_{5}$ and $q_{6}$ into Eq.~\eqref{eq:sys a b 1}, or equivalently
into Eq.~\eqref{eq:sys a b 2}.  Hence, the solutions sought for the DH
layer are \begin{eqnarray}
\dispot=\sum_{n=1,3,5,6}A_{n}e^{kq_{n}x_{2}}e^{ik\left(x_{1}-\phasespeed t\right)}, &  & \edpot=\sum_{n=1,3,5,6}kB_{n}e^{kq_{n}x_{2}}e^{ik\left(x_{1}-\phasespeed t\right)},\end{eqnarray}
where the six unknowns are  $A_{1},A_{3},A_{5},A_{6},B_{1}$ and
$B_{3}$,  while  $B_{5}$ and $B_{6}$ are functions of $A_{5}$ and
$A_{6}$.

We note that once the layer is excited the symmetry in paths A and
B is broken. Accordingly, for all paths the incremental exterior fields
must be accounted for. Motivated by the need to
satisfy a decaying condition at $x_{2}\rightarrow\pm\infty$ along
with Laplace's equation outside the material,  we consider the following
stream functions  \begin{eqnarray}
\outer{\psi} & = & C_{1}ike^{-kx_{2}}e^{ik\left(x_{1}-\phasespeed t\right)}\ \mathrm{at}\ x_{2}>h,\\
\outer{\vartheta} & = & C_{2}ike^{kx_{2}}e^{ik\left(x_{1}-\phasespeed t\right)}\ \mathrm{at}\ x_{2}<-h,\end{eqnarray}
such that the electric field components are given by \begin{eqnarray}
\outer{\inc{\efcomp{}}}_{1}=\begin{cases}
-\outer{\psi_{,1}} & x_{2}>h\\
-\outer{\vartheta_{,1}} & x_{2}<-h\end{cases}, & \quad\outer{\inc{\efcomp{}}}_{2}=\begin{cases}
-\outer{\psi_{,2}} & x_{2}>h\\
-\outer{\vartheta_{,2}} & x_{2}<-h\end{cases} & .\label{eq:outer e comp}\end{eqnarray}

In summary, we end up with a set of eight constants to be determined,
namely $\left\{ A_{n}\right\} ,\ n=1,3,5,6$,   $\left\{ B_{n}\right\} ,n\ =1,3$
and $C_{1}$, $C_{2}$ via the appropriate jump conditions. Henceforth we distinguish the solutions of paths A and B
from the one along path C.

\subsection*{Paths A and B}

The Maxwell stress, and as a consequence of Eq.~\eqref{eq:inc maxwell stress},
its increment vanish outside the layer ($\outer{\cauchy}=\inc{\outer{\cauchy}}=\mathbf{0}$).
Further, the surfaces remain traction-free and hence the jump in the
stress according to Eq.~\eqref{eq:in stress jump} becomes \begin{eqnarray}
\piolaincpushcomp{22} & = & 0\ \mathrm{at}\ x_{2}=\pm h,\label{eq:exp traction free s22}\\
\piolaincpushcomp{12} & = & 0\ \mathrm{at}\ x_{2}=\pm h,\label{eq:eq:exp traction free s12}\end{eqnarray}
where
\begin{eqnarray}
\piolaincpushcomp{22}=-\left(\dnhcc+p\right)\dispot_{,12}-\inc p-2\dnhb\edpot_{,1},
\label{eq:push of inc sigma explicit1}\\ 
\piolaincpushcomp{12}=\left(\dnhcc\right)\dispot_{,22}-p\dispot_{,11}+\dnhb\edpot_{,2}.
\label{eq:push of inc sigma explicit2}
\end{eqnarray}
The solution sought for the incremental pressure is
\begin{equation}
\inc p\left(x_{1},x_{2},t\right)=k\left(P_{1}e^{kq_{1}x_{2}}+P_{3}e^{kq_{3}x_{2}}\right)e^{ik\left(x_{1}-ct\right)},\label{eq:inc pressure explicit}\end{equation}
where $P_{1}$ and $P_{3}$ are determined via Eqs.~\eqref{eq:eom x}-\eqref{eq:eom y}.

As the electrodes surface charge is fixed we have that $\inc w_{e}=0$,
and the corresponding jump condition in Eq.~\eqref{eq:in charge  jump}
becomes \begin{eqnarray}
\outer{\inc{\edcomp{}}}_{2}-\edlincpushcomp 2 & = & 0\ \mathrm{at}\ x_{2}=\pm h,\label{eq:charge jump explicit}\end{eqnarray}
where $\outer{\inc{\edcomp{}}}_{2}=\perm\outer{\inc{\efcomp{}}}_{2}$.
The jump in Eq.~\eqref{eq:inc elec jump} is  \begin{equation}
\outer{\inc{\efcomp{}}}_{1}-\efincpushcomp 1=0\ \mathrm{at}\ x_{2}=\pm h,\label{eq:jump e explicit}\end{equation}
where $\efincpushcomp 1=\dnhb\left(\dispot_{,22}-\dispot_{,11}\right)+\frac{1}{\matper}\edpot_{,2}$.
Note that while the increment in the Maxwell stress is zero as discussed
before, Eqs.~\eqref{eq:charge jump explicit}-\eqref{eq:jump e explicit}
state that $\outer{\dot{\ed}}$ and $\outer{\dot{\ef}}$ do not vanish.

\subsection*{Path C}
Along this path the Maxwell stress
and its increment do not vanish outside the layer, yielding the following
form of Eq.~\eqref{eq:in stress jump}
\begin{eqnarray}
\piolaincpushcomp{22}-\maxinccomp{22}+\maxcomp{22}\disgradcomp{22} & = & 0\ \mathrm{at}\ x_{2}=\pm h,\label{eq:exp traction free s22 path c}\\
\piolaincpushcomp{12}-\maxinccomp{12}+\maxcomp{11}\disgradcomp{21} & = & 0\ \mathrm{at}\ x_{2}=\pm h,\label{eq:exp traction free s12 path c}\end{eqnarray}
where \begin{eqnarray}
\maxcomp{11}=-\maxcomp{22}=-\frac{\edcomp 2^{2}}{2\perm}, & \quad\maxinccomp{22}=\perm\outer{\inc{\efcomp{}}}_{2}\outer{\efcomp 2}, & \quad\maxinccomp{12}=\perm\outer{\inc{\efcomp{}}}_{1}\outer{\efcomp 2}.\label{eq:maxwell components path c}\end{eqnarray}
The expressions for $\piolaincpushcomp{22}$ and $\piolaincpushcomp{12}$
are given in Eqs.~\eqref{eq:push of inc sigma explicit1}-\eqref{eq:push of inc sigma explicit2}.
The external electric fields do not vanish in this path, and hence Eq.~\eqref{eq:in charge  jump}
assumes the form

\begin{eqnarray}
\outer{\inc{\edcomp{}}}_{2}-\edlincpushcomp 2+\outer{\edcomp 2}\disgradcomp{22} & = & 0\ \mathrm{at}\ x_{2}=\pm h,\label{eq:charge jump explicit path c}\end{eqnarray}
where $\outer{\edcomp 2}=\edcomp 2.$ The remaining jump in Eq.~\eqref{eq:inc elec jump}
is \begin{equation}
\outer{\inc{\efcomp{}}}_{1}-\efincpushcomp 1+\outer{\efcomp 2}\disgradcomp{21}=0\ \mathrm{at}\ x_{2}=\pm h.\label{eq:jump e explicit path c}\end{equation}

Eqs.~\eqref{eq:exp traction free s22}-\eqref{eq:jump e explicit}
and Eqs.~\eqref{eq:exp traction free s22 path c}-\eqref{eq:jump e explicit path c}
complete the necessary set of eight boundary conditions for the eight
unknowns for paths A and B, and path C, respectively. Furthermore,
they constitute a set of linear homogeneous equations in the unknowns,
hence non-trivial solutions exist when the determinant of the coefficients
matrix vanishes. This, in turn, is the extension of the well-known
Rayleigh-Lamb transcendental equation for a purely elastic layer,
and Tiersten's result for a piezoelectric plate \citep{Tiersten1963}
within the framework of infinitesimal deformations. Of practical interest is the
dependency of the dispersion relation, usually given in terms of the
frequency spectrum, on the bias field and the finite deformation.
These are discussed in the next section

\section{Analysis of the dispersion relation}

The working scheme of this section can be formulated as follows: \emph{Given
an angular  frequency of excitation} $\afreq$, \emph{we determine
the associated wavenumbers $\left\{ k_{m}\right\} $ satisfying the
generalized transcendental equation, determine the velocities of the
propagating waves $c_{m}=\afreq/k_{m}$, and examine how these vary
as functions of the bias field   and pre-stretch. }Specifically,
for paths A and C the dependency on the bias field is examined for
a few representative values of $\dh$. For path B, having the deformed
configuration fixed we find it advantages to explore the dependency
in terms of $\sdh$. Recalling that along this path the pre-stretch
and the electric displacement field are independent, the dependency
of the dispersion relation on a few representative values of the
pre-stretch $\tilde{\lambda}$  is also examined. Of special interest
is the identification and evaluation of the fundamental modes, i.e.,
those modes having finite velocity in the limit of long waves (e.g., \citealp{Lutianov2010}).

We find it useful to represent the results in terms of the dimensionless
quantities $\kh=hk$, $\speedh=c/\bulkspeed$ and $\afreqh=\speedh\kh$, where
$\bulkspeed=\sqrt{\mu/\rho}$ is the bulk shear wave velocity in an isotropic
elastic material. Further, our numerical investigation
implies that the value of the parameter $\relper$ has only minor
a influence on the frequency spectrum due to the choices $\dh=\edlcomp 2/\sqrt{\mu\matper}$
and $\hat{\edcomp{}}=\edcomp 2/\sqrt{\mu\matper}$ which are normalized
by the material permittivity. Accordingly, as a representative value
we chose $\matper_{r}=3$.

In their fundamental works Lord \citet{lordrayleigh89} and \citet{lamb89}
had shown that in the purely elastic case a decomposition of the mechanical
vibrations can be made into symmetric (extensional) and antisymmetric
(flexural) modes with respect to the mid-plane of the layer. Interestingly,
a similar decomposition of the mechanical displacement is possible
for the coupled problem. Interestingly, it turns out that the symmetry of
the electrical displacements is reversed in the sense that whenever
the mechanical displacements are symmetric, the electrical displacements
are antisymmetric, and vice versa. To distinguish between the two
types of modes we employ the following procedure. For  the antisymmetric
modes we subtract Eq.~\eqref{eq:exp traction free s22 path c} evaluated
at $x_{2}=-h$ from its value at $x_{2}=h$, and similarly with Eq.
\eqref{eq:jump e explicit path c}. Subsequently, we add Eq.~\eqref{eq:exp traction free s12 path c}
evaluated at $x_{2}=-h$ to its value at $x_{2}=h$, and similarly
with Eq.~\eqref{eq:charge jump explicit path c}. We end up with a
system of four homogeneous equations in terms of the four constants 
\begin{eqnarray}
\newcofa=A_{1}+A_{3},\quad\newcofaa=A_{5}+A_{6}, & \quad\newcofb=B_{3}-B_{1},\quad\newcofc=C_{1}+C_{2}.\label{eq:A+}\end{eqnarray}
The dispersion relation derived from the determinant of this system
provides the speed of propagation of waves associated with the antisymmetric
modes.
The symmetric modes can be derived by changing the foregoing
operations. Thus, to add Eq.~\eqref{eq:exp traction free s22 path c}
evaluated at $x_{2}=-h$ to its value at $x_{2}=h$, and similarly
with Eq.~\eqref{eq:jump e explicit path c}. Analogously, to subtract
Eq.~\eqref{eq:exp traction free s12 path c} evaluated at $x_{2}=-h$
from its value at $x_{2}=h$, and also with Eq.~\eqref{eq:charge jump explicit path c}.
These operations lead  to   system of four homogeneous
equations in terms of the four constants \begin{eqnarray}
A_{1}^{-}=A_{1}-A_{3},\quad A_{5}^{-}=A_{5}-A_{6}, & \quad B^{+}=B_{3}+B_{1},\quad C^{-}=C_{1}-C_{2}.\label{eq:A-}\end{eqnarray}

\subsection*{Path A: expansion-free finite loading}

\begin{figure}[t]
\includegraphics[bb=0bp 0bp 595bp 842bp,angle=-90,scale=0.5]{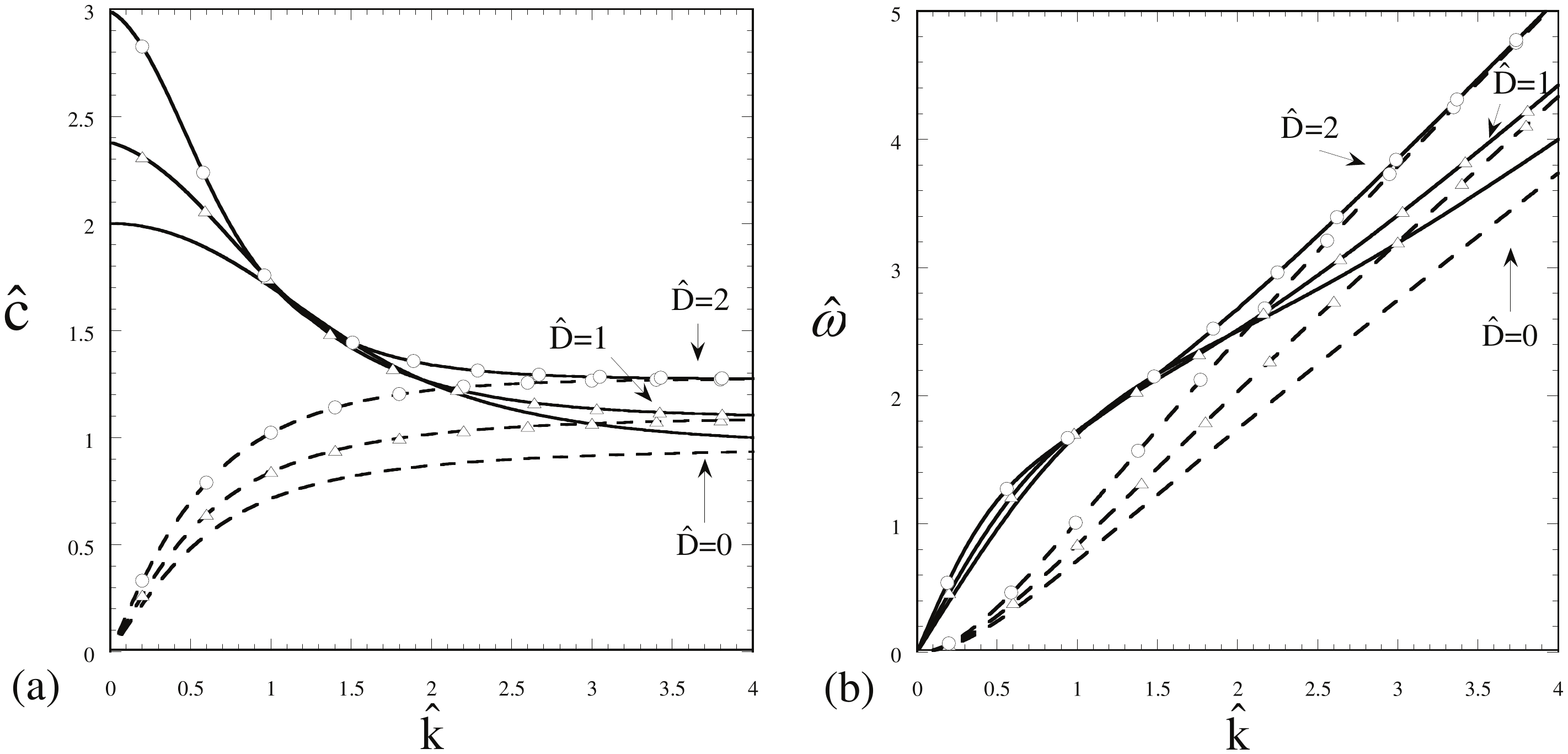}

\caption{Path A: the normalized velocity $\speedh$ (a) and angular frequency
$\afreqh$ (b) as functions of the normalized wavenumber $\kh$ for
the expansion-free finite loading path. The continuous and dashed
curves correspond to the symmetric and antisymmetric parts, respectively.
The markless curve, and the curves with triangle\textcolor{red}{{} }and
circle marks correspond to $\dh=0,1$ and 2, respectively. }

\label{dispersion a}
\end{figure}

Fig.~\ref{dispersion a}a displays the normalized velocity $\speedh$
as a function of the normalized wavenumber $\kh$ for a few values of
$\dh$. Herein and henceforth the continuous and dashed curves correspond
to the symmetric and antisymmetric modes, respectively. The markless
curve, and the curves with triangle and circle marks
correspond to $\dh=0,1$ and 2, respectively. These values of the
biased fields induce the principal stretches $\lambda=1,1.19$ and
1.49, respectively.

The case $\dh=0$ corresponds to the purely elastic problem. In this
case the generalized solution recovers the classic Rayleigh-Lamb dispersion
relation, as it should. Note that one can perceive variation in $\kh$
as variation in the plate thickness for a fixed wavenumber. Thus,
as $\kh$ increases the plate geometry becomes similar to that of
a half-space. Accordingly, the symmetric and the antisymmetric branches
become closer, such that in the limit of $\kh\rightarrow\infty$ they
coincide and attain the surface wave velocity. In particular, the
fundamental modes propagate at a velocity $\speedh=0.955$ in
the limit of short waves (large $\kh$), which is the propagation
velocity of Rayleigh surface waves. Further, the fundamental symmetric
mode attains the well known result $\speedh=2$ in the limit of long
waves, or, equivalently, thin layers (small $\kh$).

As $\dh$ increases a monotonous rise of the velocity in the limit
of long waves for the symmetric mode is revealed. The velocity at
this limit can be derived analytically upon taking the first term
in the Taylor series expansion of the symmetric dispersion relation in the
neighborhood of $\kh=0$. Equating it to zero yields an explicit expression
for the velocity in the limit of long waves, namely\begin{equation}
\speedh=2\left(1+\dh^{2}\right)^{1/4}=2\lambda.\label{eq:c long wave path a}\end{equation}
A monotonous rise of the velocity in the limit of short waves is observed
too. We stress that the asymptotic values of the dimensionless speed
are in agreement with the corresponding results obtained with an analysis
of surface waves.  

We find that all antisymmetric branches emerge at the origin, so
that no loss of stability in terms of diffuse mode is identified,   i.e.,
a vanishing $\speedh$ for non-zero wavenumbers. For
this loading path a possible instability could be macroscopic instability
associated with the loss of positive definiteness of the tangent constitutive
operator (\citealp{zhao07}, \citealp{Bertoldi2011}) that however
cannot occur for a DH layer with vanishing longitudinal stress.

Fig.~\ref{dispersion a}b displays the normalized angular frequency
$\afreqh$ as function of the normalized wavenumber $\kh$. We point out
that the slope   of the curves remains positive. Consequently, a unique
wavenumber $\kh$ is associated with each value of $\afreqh$ along
a specific curve. In other words, exciting the layer along this path
with a certain frequency will give rise to a single wavelength of
each fundamental mode, and increasing values of $\afreqh$ will
result in shorter waves.

\subsection*{Path B: Pre-stretched layer }

\begin{figure}[t]
\includegraphics[angle=-90,scale=0.5]{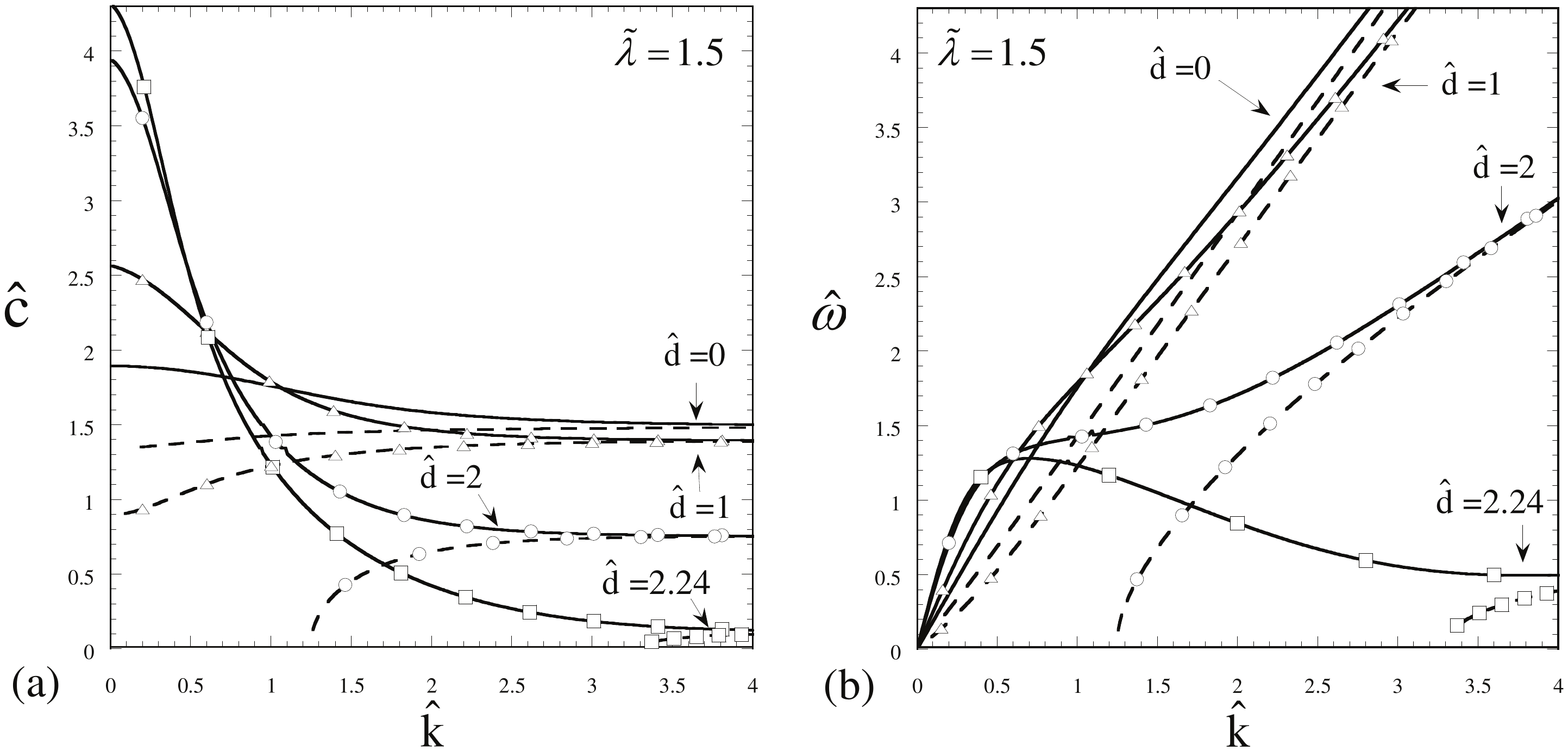}

\caption{Path B: the normalized velocity $\speedh$ (a) and angular frequency
$\afreqh$ (b) as functions of the normalized wavenumber $\kh$ for
for $\tilde{\lambda}$=1.5. The continuous and dashed curves correspond
to the symmetric and antisymmetric parts, respectively. The markless
curve, and the curves with triangle, circle and square marks correspond
to $\sdh=0,1,2$ and 2.24, respectively. }

\label{dispersion b 1.5}
\end{figure}
\begin{figure}[t]
\includegraphics[angle=-90,scale=0.5]{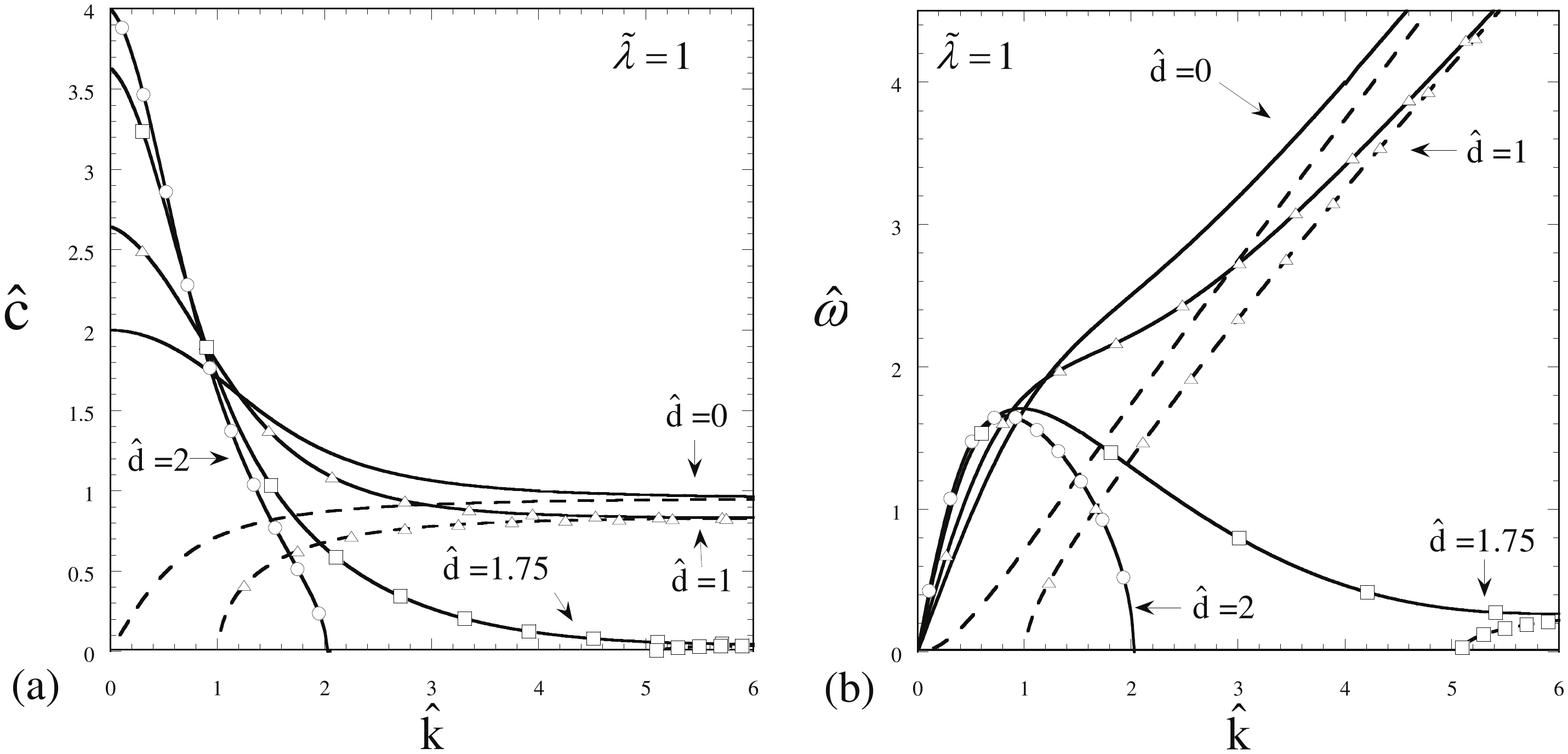}

\caption{Path B: the normalized velocity $\speedh$ (a) and angular frequency
$\afreqh$ (b) as functions of the normalized wavenumber $\kh$ for
$\tilde{\lambda}$=1. The continuous and dashed curves correspond
to the symmetric and antisymmetric parts, respectively. The markless
curve, and the curves with triangle, square and circle marks correspond
to $\sdh=0,1,1.75$ and 2, respectively. }

\label{dispersion b 1}
\end{figure}
\begin{figure}[t]
\includegraphics[angle=-90,scale=0.5]{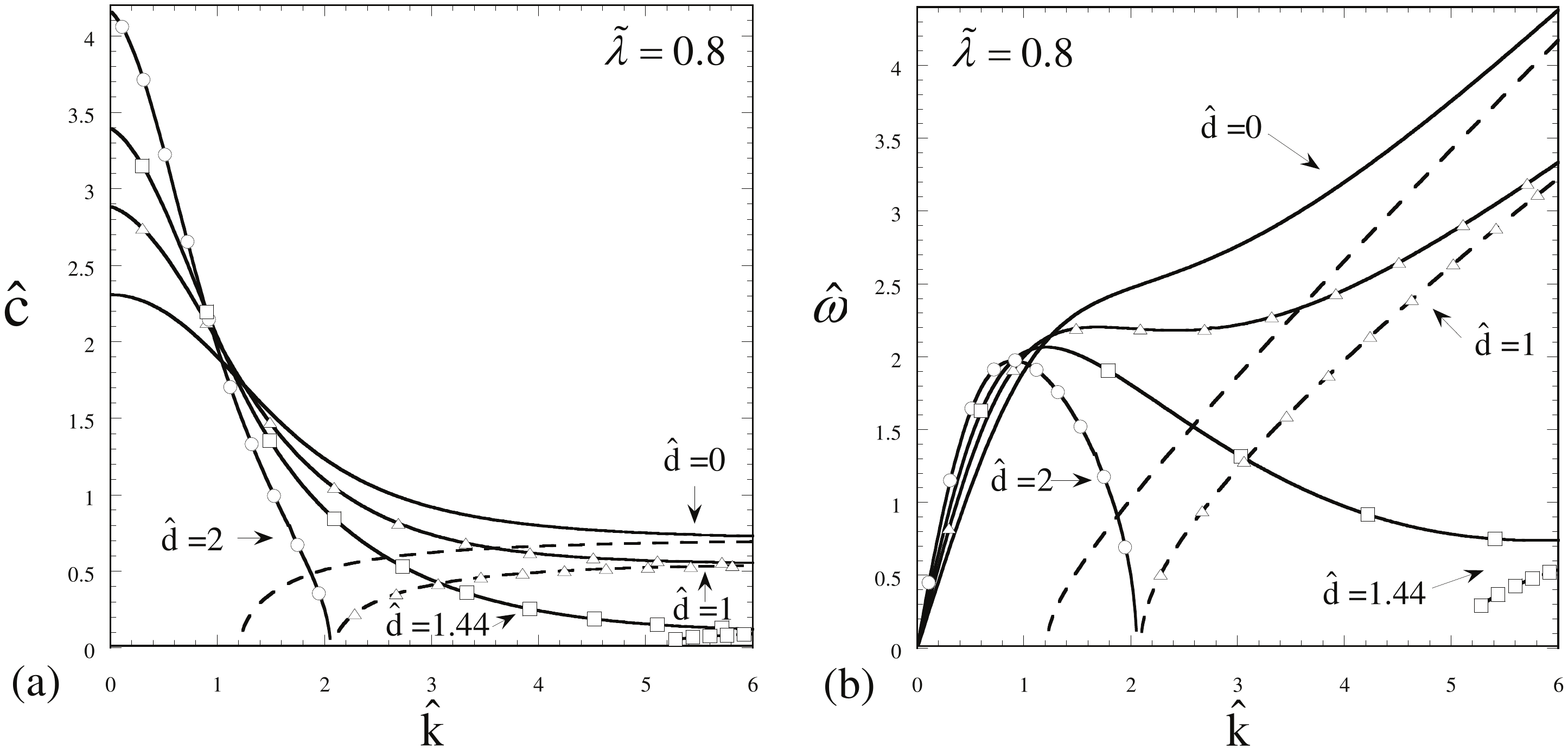}

\caption{Path B: the normalized velocity $\speedh$ (a) and angular frequency
$\afreqh$ (b) as functions of the normalized wavenumber $\kh$ for
$\tilde{\lambda}$=0.8. The continuous and dashed curves correspond
to the symmetric and antisymmetric parts, respectively. The markless
curve, and the curves with triangle, square and circle marks correspond
to $\sdh=0,1,1.44$ and 2, respectively. }

\label{dispersion b 08}
\end{figure}

Three representative types of pre-stretch where chosen, namely a tensile
pre-stretch with $\tilde{\lambda}=1.5$, no pre-stretch $\tilde{\lambda}=1$
(the layer is clamped in its the reference configuration), and a pre-compression
with $\tilde{\lambda}=0.8$. We recall that here we find it useful
to use the current quantity $\hat{d}=d_{2}/\sqrt{\mu\matper}$ as
a measure for the bias field, as $\tilde{\lambda}$ is held  fixed.
Figs. \ref{dispersion b 1.5}a, \ref{dispersion b 1}a and \ref{dispersion b 08}a
display the normalized velocity {\small $\speedh$ }\textcolor{black}{as a 
function of the normalized wavenumber $\kh$ for $\tilde{\lambda}=1.5,1$
and $0.8$, respectively. The continuous and dashed curves correspond
to the symmetric and antisymmetric parts, respectively.} The markless
curve, and the curves with triangle and circle marks correspond to
$\sdh=0,1$ and 2, respectively. The curves with square marks correspond
to $\hat{d}=2.24,1.75$ and 1.44 in Figs. \ref{dispersion b 1.5}-\ref{dispersion b 08},
respectively.

With regard to the symmetric branches, the velocity in the limit of
long waves increases with values of $\hat{d}$, in a manner similar
to the one observed for path A. Here again we take the first term
in the Taylor series expansion of the symmetric dispersion relation in the
neighborhood of $\kh=0$ and equate it to zero. The resulting explicit
expression for the velocity in this limit is \begin{equation}
\speedh=\sqrt{3\left(\sdh^{2}+\tilde{\lambda}^{-2}\right)+\tilde{\lambda}^{2}}.\label{eq:c long wave path b}\end{equation}
Interestingly, a reversed trend is revealed in the limit of short
waves, such that the surface wave velocity   decreases monotonically
with $\sdh$, in contrast  to the situation in path A. The difference
stems from the emergence of compressive stress in path B, as the layer
cannot elongate with increasing values of $\sdh$ due to the fixed
boundary. The surface wave velocity decays until it reaches a vanishing
value associated with a loss of surface stability. This   is attained
when a threshold value $\sdhth$ is applied, such that
\begin{equation}
\lim_{\kh\rightarrow\infty}\speedh=0.\label{eq:lim c 0}
\end{equation}
Thus $\sdhth$ can be perceived as the value of $\sdh$ at which a
half-space losses its stability. This threshold value depends on the
pre-stretch such that it increases monotonically with $\tilde{\lambda}$,
revealing a stabilizing effect of the pre-stretch. Specifically, for
$\tilde{\lambda}=1.5,\,1$ and 0.8 we calculate numerically the values
$\sdhth=2.245,1.753$ and 1.453, respectively. When $\sdh>\sdhth$
the velocity vanishes at a finite cutoff wavenumber which we denote
by $\kcos$.
In a way $\kcos$ defines a critical wavelength, beyond
which no propagation of waves is exhibited. Further increase of
$\sdh$ will result in smaller values of $\kcos$. For completeness, we note that our numerical
investigation shows that for the case $\hat{d}=0$ the solution recovers
Biot's result for loss of stability under compression of a neo-Hookean
half-space at a pre-compression of $\tilde{\lambda}=0.544$.

Turning to the antisymmetric branches, we observe an essentially different
evolution from the one in path A. Thus, in this case the branches
emerge from various points. We denote the wavenumber at
which the branch emerges by $\kcoa$. In a way $\kcoa$ defines a
critical wavelength beneath which there is no stable propagation of
longer waves. The value of $\kcoa$ increases when $\sdh$ is increased,
up to the threshold value $\sdhth$, at which $\kcoa$ tends to infinity.
We stress that $\sdhth$ is the one satisfying Eq.~\eqref{eq:lim c 0}
as expected, since in the limit of half-space the two branches coincide.

Figs. \ref{dispersion b 1.5}b, \ref{dispersion b 1}b, and \ref{dispersion b 08}b
display the dispersion relation in terms of the normalized frequency
$\afreqh$ as a function of the normalized wavenumber $\kh$, with identical
legend to the one previously used. It is observed how,  unlike the corresponding
curves in path A, for some values of $\sdh$ the symmetric curves
are not monotonous. In some cases the slope of the curves  is changing from positive to negative
and to positive once again. This corresponds to a situation where for
the same frequency there are more than one wavenumber   satisfying the dispersion relation for  the fundamental
symmetric mode, and subsequently
the pertained different wavelengths propagate with different velocities.
If the frequency is to be taken as the independent variable, this suggests a peculiar phenomenon in which
as the frequency is increased there is a possibility for the appearance of shorter and slower waves that propagate along with the 
``primary'' waves.

\subsection*{Path C: A layer immersed in an electric fields}

\begin{figure}[t]
\includegraphics[angle=-90,scale=0.5]{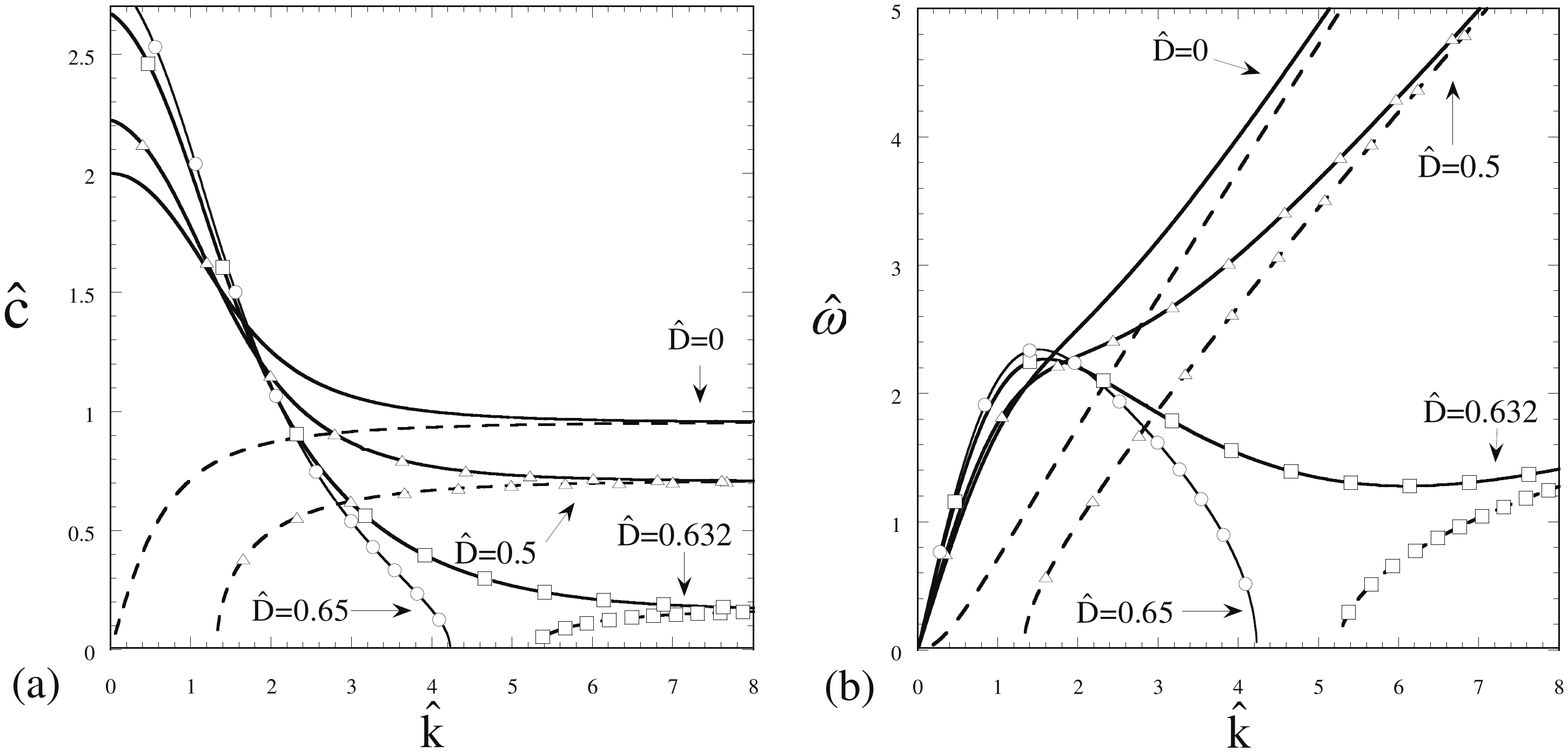}

\caption{Path C: the normalized velocity $\speedh$ (a) and angular frequency
$\afreqh$ (b) as functions of the normalized wavenumber $\kh$ for
a layer immersed in an electric field. The continuous and dashed curves
correspond to the symmetric and antisymmetric parts, respectively.
The markless curve, and the curves with triangle, circle and square
marks correspond to $\dh=0,0.5,0.632$ and 0.65, respectively. }

\label{dispersion c}
\end{figure}

We recall that Eq.~(\ref{eq: lambda d third path}a) states that the
layer   expands along the direction of the electric displacement,
in a manner opposite to the one revealed in paths A and B. Further,
the stretch depends strongly on the material permittivity, even when
the normalized quantity $\dh$ is used. For the choice $\relper=3$
the maximal value of admissible $\dh$ is $1/\sqrt{2}$, as $\lambda$
vanishes for this value in view of Eq. (\ref{eq: lambda d third path}a).
Consequently, the values of $\dh$ considered  in path A cannot be
used, and even values of $\dh$ that result in the same values of
$\lambda$ are not feasible as here $\lambda<1$, whereas along path
A $\lambda>1$. Nonetheless, attempting  to investigate cases analogue
to those addressed in path A,  we choose values of $\dh$  that  yield
values of $\max\left\{ \lambda,1/\lambda\right\} $ similar to those
considered in path A.

Fig.~\ref{dispersion c}a displays the normalized velocity $\speedh$
as a function of the normalized wavenumber $\kh$ for $\dh=0,0.5,0.632$
and 0.65. These values correspond to principle maximal stretche ratios
$1,1.2,1.5$ and 1.59, respectively. The continuous and dashed curves
correspond to the symmetric and antisymmetric solutions, respectively.
The markless curve, and the curves with triangle, circle and square
marks correspond to $\dh=0,0.5,0.632$ and 0.65, respectively. Once
again, the symmetric modes exhibit a rise in the velocity in the limit
of long waves with increasing values of $\dh$. In a manner similar to  path B, the trend
is reversed in the short waves limit, and the surface wave velocity
decreases monotonically as $\dh$ increases. This   is a consequence
of the resultant compressive stress that can be determined   by substitution
of Eq.~(\ref{eq:maxwell components path c}a) into Eq.~(\ref{eq:jumps}a)
for the jump in the stress. Thus, the surface wave velocity decays as
a function of $\dh$, until it reaches a vanishing value associated
with a loss of surface stability, i.e., satisfying Eq.~\eqref{eq:lim c 0}.
Once again  we denote the threshold value of $\dh$ for which
the latter holds by $\dhth$. In this numerical example we find that
$\dhth=0.636$. As in Path B, when $\dh>\dhth$ the velocity of the
symmetric modes vanishes at a finite cutoff wavenumber $\kcos$ beyond
which no shorter waves are exhibited.

The behavior of the antisymmetric modes is also reminiscent of the one observed for
path B. Specifically, it is observed how the curves emerge at different
cutoff wavenumbers $\kcoa$ defining critical wavelength beneath which
there is no stable propagation of longer waves. The cutoff wavenumber
increases monotonically with values of $\dh$, and reaches infinity
when $\dh$ attains the aforementioned value $\dhth$. Subsequently,
when $\dh>\dhth$ there is no stable antisymmetric wave propagation.

Fig.~\ref{dispersion c}b displays the dispersion relation in terms
of the normalized frequency $\afreqh$ as function of the normalized
wavenumber $\kh$. Trends similar to the ones observed in path B are
revealed. In particular, the monotonicity is lost when $\dh>0.584$,
where the positive slop of the curve becomes negative and then positive
again. Here again, this implies that imposing certain frequencies may yield
more than one wavelength of the fundamental symmetric mode satisfying
the dispersion relation. In turn, this suggests that when the frequency is increased there is a possibility of simultaneous different phase velocities.

\section{Concluding remarks}
In view of the emergence of modern  EAPs capable of large deformations  in response to electric excitation we examine the  topic of small waves propagation superposed on finitely deformed dielectric elastomer layers. Specifically,  we considered a layer whose electromechanical behavior is characterized by the DH model when subjected to a coaxial finite deformation. Three different loading paths which result in the aforementioned finite configurations were addressed. In path A the layer is free to expand along its longitudinal direction and deformed due to the  electric excitation  along its thickness. In path B the layer is initially pre-stretched, then clamped, and subsequently the electric field is employed. In path C the layer is  immersed in a pre-existing electric field.

Following \citet{Dorfmannogden2010} small perturbations on top of the homogeneous finite deformation are considered. The   coupled equations of incremental electrodynamics along with the pertained boundary conditions yielded the desired dispersion relation. For the DH model this required a non-conventional presentation of the solution of the governing equations.  It was shown that a separation of the mechanical and electric waves to   symmetric and antisymmetric modes with respect to the mid-plane of the layer is feasible. We find that the symmetric mechanical waves are accompanied by antisymmetric electric waves and vice versa.
 
 Numerical investigation of the fundamental modes was conducted  to examine the influence of the bias field along  the three loading paths.  For   path B  we also examined the effect of pre-stretch. In path A the velocity of both short waves (half-space, large $\kh$) and long waves (thin plate, small $\kh$) is increased when the   electric displacement field is enhanced. While in the limit of long waves  a similar rise  in the velocity is observed in path B, the velocity decreases    as function of the   electric displacement field in the limit of short waves. The existence of a threshold values of $\sdh$  at  which  the layer losses stability  is observed. The pre-stretch effect is shown to be stabilizing, as higher values of $\sdh$ are needed to reach the onset of  instability. Dispersion relations similar to the ones obtained in path B were observed in path C. We finally stress that along all paths there is a marked  influence of the bias electric field and pre-stretch on the propagation speed. This phenomenon lends itself as a possible control mechanism for the speed of the waves, or even for filtering specific wavelengths with suitable adjustment of the bias field and pre-stretch.

\section*{Acknowledgements}

The first author wishes to thank a scholarship in the framework of the Erasmus Mundus External Cooperation Window (grant lot3.emecw.com). The second author gratefully acknowledges the financial support of COST Action MP1003 "European Scientific Network for Artificial Muscles".


\begin{small}

\end{small}

\end{document}